\documentclass{article}
\usepackage{spconf,graphicx,hyperref,subcaption}
\usepackage{xcolor}
\usepackage{algorithm}
\usepackage{algorithmic}
\usepackage{amsmath,amssymb,amsfonts}

\usepackage[official]{eurosym}
\usepackage{theorem}
\usepackage[usenames,dvipsnames]{pstricks}
\usepackage{euscript}
\usepackage{charter}
\usepackage{comment}
\usepackage{array,multirow}
\usepackage{float}

\input{alphabet}

\newcommand{\argmind}[1]{\ensuremath{\underset{\substack{{#1}}}%
{\mathrm{argmin}}\;\; }}
 
\newcommand*{\prox}{{\mathrm{prox}}}

\title{Efficient Plug-and-Play Method for Dynamic Imaging via Kalman Smoothing}
\name{Benjamin Hawkes $^{\ast \star \dagger}$, Mike Davies $^{\dagger}$, V\'ictor Elvira $^{\ddagger\diamond}$, Audrey Repetti $^{\ast \diamond}$ \thanks{This work was supported by the EPSRC and MoD SPADS CDT [EP/Y013859/1]. The work of AR was partly supported by the EPSRC grant [EP/X028860/1].}}

\address{ \small
$^\ast$ School of Mathematical and Computer Sciences, Heriot-Watt University, Edinburgh, United Kingdom \\ 
\small 
$^\star$ School of Engineering and Physical Science, Heriot-Watt University, Edinburgh, United Kingdom \\ \small
$^\dagger$ Institute of Digital Communications, University of Edinburgh, Edinburgh, United Kingdom \\ \small
$^\ddagger$ School of Mathematics, University of Edinburgh, Edinburgh, United Kingdom \\ 
\small 
$^\diamond$ Maxwell Institute for Mathematical Sciences, Edinburgh, United Kingdom}
\begin{document}
%
\maketitle
\begin{abstract}

State-space models (SSM) are common in signal processing, where Kalman smoothing (KS) methods are state-of-the-art. However, traditional KS techniques lack expressivity as they do not incorporate spatial prior information. Recently, \cite{gao2019iterated} proposed an ADMM algorithm that handles the state-space fidelity term with KS while regularizing the object via a sparsity-based prior with proximity operators. 
Plug-and-Play (PnP) methods are a popular type of iterative algorithms that replace proximal operators encoding prior knowledge with powerful denoisers such as deep neural networks. 
These methods are widely used in image processing, achieving state-of-the-art results.
In this work, we build on the KS-ADMM method, combining it with deep learning to achieve higher expressivity.
We propose a PnP algorithm based on KS-ADMM iterations, efficiently handling the SSM through KS, while enabling the use of powerful denoising networks.
Simulations on a 2D$+$t imaging problem show that the proposed PnP-KS-ADMM algorithm improves the computational efficiency over standard PnP-ADMM for large numbers of timesteps.
\end{abstract}
%
\section{Introduction}
\label{sec:intro}

Inverse imaging problems consist of inferring an unknown image from noisy measurements. In this work, we assume that at time $t\in \{1, \ldots, T\}$, for $T\in \eN$, an unknown image $\overline{x}_t\in \eR^N$ is observed through a linear model given by 
\begin{equation}\label{eq:inv-prob}
    y_t = H_t \overline{x}_t + r_t
\end{equation}
where $y_t \in \eR^M$ is the measurement, $H_t \in \eR^{M \times N}$ is the linear measurement operator, and $r_t \in \eR^M$ is a realization of a white Gaussian random variable with covariance $R_t \in \eR^{M \times M}$. 
Inverse problems are widespread in signal processing, with applications from astronomy to medicine and beyond. We further consider dynamic inverse problems where the unknown images are linked together through a linear system given by
\begin{equation}\label{eq:state}
(\forall t \in \{2, \ldots, T\})\quad
    \overline{x}_t = A_t \overline{x}_{t-1} + q_t,
\end{equation}
where $A_t \in \eR^{N \times N}$ is the linear state transition operator and $q_t \in \eR^N$ is a realization of a white Gaussian random variable with covariance $Q_t \in \eR^{N \times N}$. 
The resulting framework~\eqref{eq:inv-prob}-\eqref{eq:state} is often referred to as a linear Gaussian state space model (SSM) when $T>1$ \cite{sarkka2023bayesian}.
In imaging, SSMs have been used to model temporal evolution across frames to perform different video processing tasks such as super-resolution \cite{feng2024kalman} and denoising \cite{arias2019kalman}.

In the case of $T=1$, 
one can define the estimate $\widehat{x}_t$ of $\overline{x}_t$ to be a minimizer of a regularized least-squares problem. Recently, PnP methods have gained popularity in image restoration for achieving high-quality reconstructions by combining iterative proximal methods \cite{combettes2011proximal} with deep learning. Usually, the proximal operator encoding the regularization is replaced by a powerful denoiser \cite{kamilov2023plug}. This approach has been used within different proximal schemes, including ADMM \cite{reehorst2018regularization, cohen2021regularization}, half-quadratic splitting \cite{zhang2021plug}, forward-backward \cite{kamilov2017plug, pesquet2021learning} or primal-dual algorithms \cite{garcia2023primal}. 
In particular, \cite{ryu2019plug,hurault2022proximal} provide theoretical guarantees for the ADMM-PnP for suitable choices of denoisers.
Further, \cite{monod2025videopnp} recently proposed a PnP-ADMM method for when $T>1$ using a video denoiser (processing the full video), only handling~\eqref{eq:inv-prob}, and discarding the dynamical state model~\eqref{eq:state}. 

For more generic cases where $T>1$, Kalman-based techniques are state-of-the-art \cite{sarkka2023bayesian}, but they usually do not use additional prior knowledge on the target images $(\overline{x}_t)_{1 \le t \le T}$ (e.g., sparsity or smoothness, possibly in a transformed domain such as wavelets). To overcome this, hybrid methods mixing Kalman-based techniques and proximal-based algorithms have appeared recently \cite{akyildiz2018incremental, akyildiz2019probabilistic}. 
In particular, the authors in \cite{gao2019iterated} consider an ADMM algorithm, where a Kalman smoother (KS) and an iterated extended KS are used to implement steps of the ADMM associated with the SSM. 
The resulting KS-ADMM has shown to be more computationally efficient than standard ADMM as the state estimation for multiple frames is less computationally expensive.

In this work, we propose a PnP KS-ADMM approach, where, at each iteration, a KS step handling the SSM is combined with a denoising network. The resulting algorithm efficiently couples the frames through smoothing recursions rather than large-batch ADMM solves, while achieving high-quality reconstruction via powerful PnP denoisers.
We will show through simulations that the resulting PnP KS-ADMM enables PnP-type quality image reconstructions, whilst enabling significant speeding up of the reconstruction for long time sequences.

The remainder of the paper is structured as follows. Section~\ref{sec:background} introduces relevant background. Section~\ref{sec:PropModel} outlines the proposed model. Section~\ref{sec:VidDeblur}, gives simulation results on a synthetic $2D+t$ deblurring task and we show its computational benefits over a standard PnP-ADMM algorithm.  We conclude in Section~\ref{sec:Conc}.

\section{Background}
\label{sec:background}

We denote sequences of $T$ variables with bold, i.e., $\xb = (x_t)_{1 \le t \le T} \in \eR^{N \times T}$. Similarly, with a slight abuse of notation, for a collection of linear operators $(H_t)_{1 \le t \le T}$ in $\eR^{M \times N}$ we denote $\Hb \xb = (H_tx_t)_{1 \le t \le T} \in \eR^{M \times T}$.

Our goal is to find an estimate $\widehat{\xb}$ of $\overline{\xb}$ from the noisy measurements $\yb$ obtained through~\eqref{eq:inv-prob}-\eqref{eq:state}. 
Since the measurement and transition operators $(H_t)_{1 \le t \le T}$ and $(A_t)_{2 \le t \le T}$ are assumed linear, a KS within a maximum likelihood framework can be used \cite{rauch1965maximum}. 

\subsection{Maximum \textit{a Posteriori} framework}

We adopt a maximum \textit{a posteriori} approach, and define the estimate $\widehat{\xb}$ as

\vspace{-1mm}
\begin{equation}\label{eq:min}
    \widehat{\xb} = 
    \argmind{\xb} F(\xb) + G(\xb),
    \vspace{-2mm}
\end{equation}
where $F \colon \eR^{N \times T} \to \eR$ is the data fidelity term associated with the SSM \eqref{eq:inv-prob}-\eqref{eq:state}, and $G \colon \eR^{N \times T} \to (-\infty, +\infty]$ is a regularization term incorporating \textit{prior} information we have on the target images $\overline{\xb}$.
In particular, $F$ can be chosen to be the least-squares criterion
\begin{equation*}
    F(\xb) = 
    \frac{1}{2}\sum_{t=1}^{T} \|y_t - H_t x_t\|^2_{R_t^{-1}} 
    + \frac{1}{2}\sum_{t=2}^{T} \|x_t - A_t x_{t-1}\|^2_{Q_t^{-1}} .
\end{equation*}
For $G$, we choose a standard regularization given by 
\begin{equation}\label{eq:G}
    G(\xb) = \frac{1}{2} \|x_1 - m_1\|^2_{P_1^{-1}} + \sum_{t=1}^{T} g_t(W_t x_t)
\end{equation}
where $m_1 \in \eR^N$ is a \textit{prior} average image for $\overline{x}_1$ and $P_1 \in \eR^{N\times N}$ a symmetric positive definite matrix. If $m_1 = \zerob_N$ and $P_1 $ is the identity matrix, then the first term in \eqref{eq:G} corresponds to a Tikhonov regularization (also known as ridge regression). In addition, for every $t \in \{1, \ldots, T\}$, $g_t \colon \eR^S \to (-\infty, +\infty]$ models a spatial regularization for each image, in the transformed domain induced by $W_t \colon \eR^N \to \eR^S$. Standard choices for $(g_t)_{1 \le t \le T}$ are the $\ell_1$ or $\ell_2$ norms, possibly combined with a wavelet transform or a total variation gradient operator encoded by $(W_t)_{1 \le t \le T}$ \cite{Chambolle_A_2016_an}.



\subsection{ADMM and KS-ADMM iterations}

A standard method for solving~\eqref{eq:min} consists in using the ADMM algorithm. In this context, Problem~\eqref{eq:min} is associated with an augmented Lagrangian \cite{parikh2014proximal, komodakis2015playing}, given by
\begin{multline}
\Lc(\xb, \wb, \etab)
    = F(\xb) + 
    \frac{1}{2} \|x_1 - m_1\|^2_{P_1^{-1}} 
        +  \sum_{t=1}^T g_t(w_t) \\
    + \etab^\top (\wb - \Wb \xb)
        + \frac{\rho}{2}  \|\wb - \Wb \xb\|^{2}.
\end{multline}
where $\wb \in \eR^{S \times T}$ is an auxiliary variable induced by $(W_t)_{1 \le t \le T}$, $\rho>0$ is a regularization parameter to enforce similarity between $\wb$ and $(W_t x_t)_{1\le t \le T}$, and $\etab \in \eR^{S \times T}$ is the dual variable (also known as the Lagrange multiplier). 
Then, an iteration $k\in \eN$ of the ADMM algorithm can be written as
\vspace{-2mm}
\begin{align}
    \xb^{(k+1)} &= \argmind{\xb} \Lc(\xb, \wb^{(k)}, \etab^{(k)}), \label{x-update} \\
    \wb^{(k+1)} &= \argmind{\wb} \Lc(\xb^{(k+1)}, \wb, \etab^{(k)}), \label{w^k+1} \\
    \etab^{(k+1)} &= \etab^{(k)} + \rho (\wb^{(k+1)} - \Wb \xb^{(k+1)}). \label{11}
\end{align}
By denoting $\Sigmab^{-1} = \rho \Ib$ and defining, for every $\xb \in \eR^{N \times T}$ and $\zb \in \eR^{S\times T}$,
\vspace{-2mm}
\begin{equation}
    \varphi(\xb, \zb) \!
       = \! F(\xb) \!
        +\!  \frac{1}{2} \|x_1 - m_1\|^2_{P_1^{-1}} 
        \!+\! \frac{1}{2} \| \Wb \xb - \zb \|^2_{\Sigmab^{-1}},
        \label{admm:x-phi} 
\end{equation}
step~\eqref{x-update} can be rewritten as
\begin{align}
    &   \zb_{\rho}^{(k)} = \rho^{-1} \etab^{(k)} + \wb^{(k)}    \label{admm:z}  \\
    &   \xb^{(k+1)} = \argmind{\xb} \varphi(\xb,  \zb_\rho^{(k)}) \label{admm:x} 
\end{align}
\vspace{-1mm}
and step~\eqref{w^k+1} can be rewritten as
\vspace{-1mm}
\begin{align}
    \wb^{(k+1)} 
    &   = \argmind{\wb} \sum_{t=1}^T g_t(w_t) \nonumber \\
    &   \qquad \qquad + \frac{\rho}{2} \|  \Wb \xb^{(k+1)} - \wb + \rho^{-1} \etab^{(k)}\|^2, \nonumber \\
    &   = \Big( \prox_{g_t} ( W_t x_t^{(k+1)} + \rho^{-1} \eta_t^{(k)}) \Big)_{1 \le  t \le T}
        \label{admm:w} 
\end{align}
\vspace{-1mm}
where $\prox$ denotes the proximity operator \cite{combettes2011proximal, moreau1965proximite}. 


\subsubsection{\textbf{ADMM computation of $\xb^{(k+1)}$}}\label{sssec:admm:admm}
Problem~\eqref{admm:x} admits a closed-form solution given by
\begin{multline}\label{x-exact}
\xb^{(k+1)}
=   \big( \Hb^\top \Rb^{-1} \Hb + \Psib^\top \Qb^{-1} \Psib 
    + \rho \Wb^\top \Wb \big)^{-1} \\
    \big( \Hb^\top \Rb^{-1} \yb 
    + \Psib^\top \Qb^{-1} \mb 
    + \rho \Wb^\top (\wb^{(k)} 
    + \rho^{-1} \etab^{(k)} ) \big)
\end{multline}
where $\mb = (m_1, 0, \ldots, 0)\in \eR^{N \times T}$ and 
$\Psib \xb = (x_1, x_2- A_2 x_1, \ldots, x_T - A_T x_{T-1})$.

When $T$ and $N$ are large, computing $\xb^{(k+1)}$ according to~\eqref{x-exact} becomes untractable. To overcome this issue, a solution to~\eqref{admm:x} can be approximated using a gradient descent (GD) or conjugate gradient (CG) sub-iterations. 
In practice, doing only one iteration of either the GD or CG has also good asymptotic behavior, while reducing drastically the computation load \cite{parikh2014proximal}. In this case, the GD iteration to approximate \eqref{admm:x} is given by
\begin{equation}\label{admm:x-GD}
    \xb^{(k+1)} = \xb^{(k)} - \nabla_{\xb} \varphi(\xb^{(k)}, \wb^{(k)}, \zb_\rho^{(k)}) .
\end{equation}
Similarly, the CG to approximate \eqref{admm:x} is given by
\begin{equation}\label{admm:x-CG}
    \xb^{(k+1)} = \xb^{(k)} - \alpha_k \, \nabla_{\xb} \varphi(\xb^{(k)}, \wb^{(k)}, \zb_\rho^{(k)}), 
\end{equation}
where $\alpha_k = (\pb_k^\top \pb_k)/(\pb_k^\top \nabla^2_{\xb} \varphi(\xb^{(k)}, \wb^{(k)}, \zb_\rho^{(k)}) \pb_k)$ and $\pb_k = - \nabla_{\xb} \varphi(\xb^{(k)}, \wb^{(k)}, \zb_\rho^{(k)})$.

\subsubsection{\textbf{KS-ADMM computation of $\xb^{(k+1)}$}}\label{ssec:admm:ks-admm}
The objective function $\varphi$ given in \eqref{admm:x-phi} appears to be the same that is optimized when using a KS for solving~\eqref{eq:inv-prob}-\eqref{eq:state}. 
Hence, instead of using either~\eqref{x-exact} or \eqref{admm:x-GD}, a KS can be used to compute $\xb^{(k+1)}$ within the ADMM iterations, as proposed in \cite{gao2019iterated}. This KS can be decomposed in four steps, described below. For simplicity, we here drop the dependency on iterations $k$ apart for the variables $(x_t^{(k)})_{1 \le t \le T}$ and $(z_{\rho, t}^{(k)})_{1 \le t \le T}$ (introduced in \eqref{admm:z}).



We first compute the prediction and update steps, going forward in time for $t\in \{2, \ldots, T\}$, i.e.,
\begin{enumerate}
    \item \textit{Prediction step:}
    \vspace{-2mm}
    \begin{align*}
        m_{\Bar{t}} &= A_t m_{t-1} \label{KS_1}\\
        P_{\Bar{t}} &= A_t P_{t-1} A_t^\top + Q_t
    \end{align*}
    where $m_{\Bar{t}}$ and $P_{\Bar{t}}$ are the predicted values of the mean and covariance at time $t$ using the dynamic model and $m_{t-1}$ and $P_{t-1}$ is the result of the forward filtering recursion at timestep $t-1$. 

    \item \textit{Update step for $y_t$:}
    \vspace{-2mm}
    \begin{align*}
        S_{y, t} &= H_t P_{\Bar{t}} H_t^\top + R_t\\
        K_{y, t} &= P_{\Bar{t}} H_t^\top S_{y, t}^{-1}\\
        m_{y, t} &= m_{\Bar{t}} + K_{y, t}(y_t - H_tm_{\Bar{t}}) \\
        P_{y, t} &= P_{\Bar{t}} - K_{y, t} S_{y, t} K_{y, t}^\top
    \end{align*}
    where $S_{y, t}$ is the innovation covariance for $y_t$, $K_{y, t}$ is the Kalman gain for $y_t$ and $m_{y, t}$ and $P_{y, t}$ are the mean and covariance when $y_t$ is considered. 

    \vspace{-1mm}
    \item \textit{Update step for $z_{\rho,t}^{(k)} = \rho^{-1} \eta_t^{(k)} + w_t^{(k)}$:}
    \vspace{-2mm}
    \begin{align*}
        S_{z, t} &= W_t P_{y, t} W_t^\top + {\Sigma}\\
        K_{z, t} &= P_{y, t} W_t^\top S_{z, t}^{-1}\\
        m_{t} &= m_{y, t} + K_{z, t}({z_{\rho,t}^{(k)}} - W_t m_{y, t}) \\
        P_{t} &= P_{y, t} - K_{z, t} S_{z, t} K_{z, t}^\top
    \end{align*}
    where $S_{z, t}$ and $K_{z, t}$ are the innovation covariance and the Kalman gain for $z_{\rho,t}^{(k)}$, respectively, {$\Sigma = \rho^{-1} \Ib$} and $m_{z, t}$ and $P_{z, t}$ are the image mean and covariance when $z_{\rho, t}^{(k)}$ is considered. 
\end{enumerate}
Then, the final step is a smoothing backward recursion, i.e., for $t \in \{T-1, \ldots, 1\}$.
\begin{enumerate}
    \setcounter{enumi}{3}
    \vspace{-1mm}
    \item \textit{Smoothing step:}
    We initialize with $x_T^{(k+1)} = m_{T}$ and $P_{s, T} = P_{T}$, and compute
    \vspace{-2mm}
    \begin{align*}
        G_t &= P_{t} A_t P_{\overline{t+1}}^{-1} \\
        x_t^{(k+1)} &= m_{t} + G_t(x_{t+1}^{(k+1)} - m_{\overline{{t}+1}})\\
        P_{s, t} &= P_{t} + G_t(P_{s, t+1} - P_{\Bar{t}+1})G_t^\top 
        \vspace{-2mm}
    \end{align*} 
    where $G_t$ is the smoothing gain, and $P_{s, t}$ is the covariance associated with $x_t^{(k+1)}$. 
\end{enumerate}

Since the Kalman smoother solves the same problem as Equation \ref{x-exact}, but leverages the dynamics of the problem, it has been shown to be more computationally efficient when working with a large number of timesteps \cite{gao2019iterated}. Further, the output of the \textit{smoothing step} also quantifies the uncertainty of $\xb^{(k+1)}$, to some extent, via the definition of its covariance matrices $(P_{s,t})_{1 \le t \le T}$.

\section{Proposed PnP KS-ADMM approach}
\label{sec:PropModel}

We propose a new algorithm, PnP KS-ADMM, including a powerful denoiser as a prior for the SSM. We replace the step handling the regularization in ADMM (see \eqref{admm:w}) by a denoising network. In this context, for every $t\in \{1, \ldots, T\}$, $W_t = \operatorname{I}$ in the previous section (so $S=N$), and the update of $\wb^{(k+1)}$ in \eqref{admm:w} is replaced by $\wb^{(k+1)} 
    = \Big( \Dc(x^{(k+1)}_t + \rho^{-1} \eta^{(k)}_t ) \Big)_{1 \le t \le T}$
where $\Dc \colon \eR^N \to \eR^N$ is a pre-trained denoising network. 
Then, the resulting PnP KS-ADMM iterations are summarized in Algorithm~\ref{algo:pnp-ks-admm}.

\begin{algorithm}[t]\small
\caption{\label{algo:pnp-ks-admm}
PnP KS-ADMM algorithm}
\begin{algorithmic}[1] 
\STATE \textbf{Set:}
$\xb^{(0)} \in \eR^{N \times T}$, $\wb^{(0)} \in \eR^{N \times T}$, $\etab^{(0)} \in \eR^{N \times T}$
\FOR{$k=0, 1, \ldots$}
\STATE Compute $\xb^{(k+1)}$ using KS given in Section~\ref{ssec:admm:ks-admm}
\STATE $\wb^{(k+1)} = \Big( \Dc(x^{(k+1)}_t + \rho^{-1} \eta^{(k)}_t) \Big)_{1 \le t \le T}$
\STATE $\eta^{(k+1)} = \etab^{(k)} + \rho (\wb^{(k+1)} -  \xb^{(k+1)})$
\ENDFOR
\end{algorithmic}
\end{algorithm}

\section{Simulation results}
\label{sec:VidDeblur}

We apply the proposed PnP KS-ADMM to a synthetic $2D+t$ problem, and we aim to compare the results with a PnP ADMM method where the image updates are computed with the strategies described in Section~\ref{sssec:admm:admm}. In particular, for the PnP ADMM, we consider the cases where either the $\xb$ update is computed exactly as in \eqref{x-exact}, with a GD sub-iteration as in \eqref{admm:x-GD} or a CG sub-iteration as in \eqref{admm:x-CG}.
We compare the different approaches when either the number of frames $T$ increases, or the image dimension increases. In all our experiments, we use a DRUNet from Deep Inverse \cite{zhang2021plug, tachella2025deepinverse}. 
Methods are implemented with \href{https://deepinv.github.io/deepinv/}{DeepInv}, and run on NVIDIA GeForce RTX 2080 Ti.

\begin{figure}[b] 
    \vspace{-4mm}
    \centering
    \setlength\tabcolsep{0.12cm}
    \small
    \begin{tabular}{ccccc}
        $t=1$ & $t=4$ & $t=7$ & $t=10$ & $t=13$ \\
       \includegraphics[width=0.13\columnwidth]{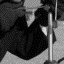}  
       &    \includegraphics[width=0.13\columnwidth]{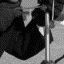}  
       &    \includegraphics[width=0.13\columnwidth]{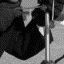}
       &    \includegraphics[width=0.13\columnwidth]{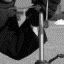}
       &    \includegraphics[width=0.13\columnwidth]{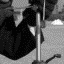} \\
       \includegraphics[width=0.13\columnwidth]{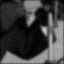}  
       &    \includegraphics[width=0.13\columnwidth]{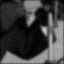}  
       &    \includegraphics[width=0.13\columnwidth]{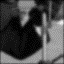}
       &    \includegraphics[width=0.13\columnwidth]{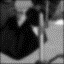}
       &    \includegraphics[width=0.13\columnwidth]{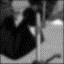} \\
        \includegraphics[width=0.13\columnwidth]{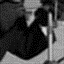} 
        &    \includegraphics[width=0.13\columnwidth]{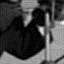}  
       &    \includegraphics[width=0.13\columnwidth]{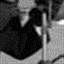}
       &    \includegraphics[width=0.13\columnwidth]{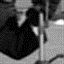}
       &    \includegraphics[width=0.13\columnwidth]{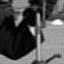} \\
       
    \end{tabular}
    \vspace{-3mm}

    \caption{\small
    Ground truth images for frames $t\in \{1, 4, 7, 10, 13\}$ (top row), with associated measurements (middle row, average PSNR~$24.05$dB), and reconstruction obtained with PnP KS-ADMM (bottom row, average PSNR~$28.55$dB).
    }
    \label{fig:recon} 
    \vspace{-2mm}
    \end{figure}

We consider a simple state-space deblurring model of the form of \eqref{eq:inv-prob}-\eqref{eq:state}, where, for every $t\in \{1, \ldots, T\}$, $H_t$ models a $2D$ convolution operator using a shared Gaussian kernel in Toeplitz form \cite{hansen2002deconvolution}. 
For the images $(x_t)_{1 \le t \le T}$, we consider a $2D$ image from which we select a patch of size $N = N_x \times N_y$, and we shift the patch $T$ times across the image. We consider patches of dimension $N \in \{32\times 32, 48 \times 48, 64 \times 64, 96\times 96\}$ and $T \in \{3, \ldots, 20\}$ frames.
For the dynamic model, as $A_t$ is unknown, we choose the identity operator in our reconstruction processes.
An example of the frames for $t\in \{1, 4, 7, 10, 13\}$ used in our simulation is shown in the top row and the associated noisy measurements are shown in the middle row of Figure \ref{fig:recon}.


\begin{figure}[t] 
    \centering
    \includegraphics[width=0.38\textwidth]{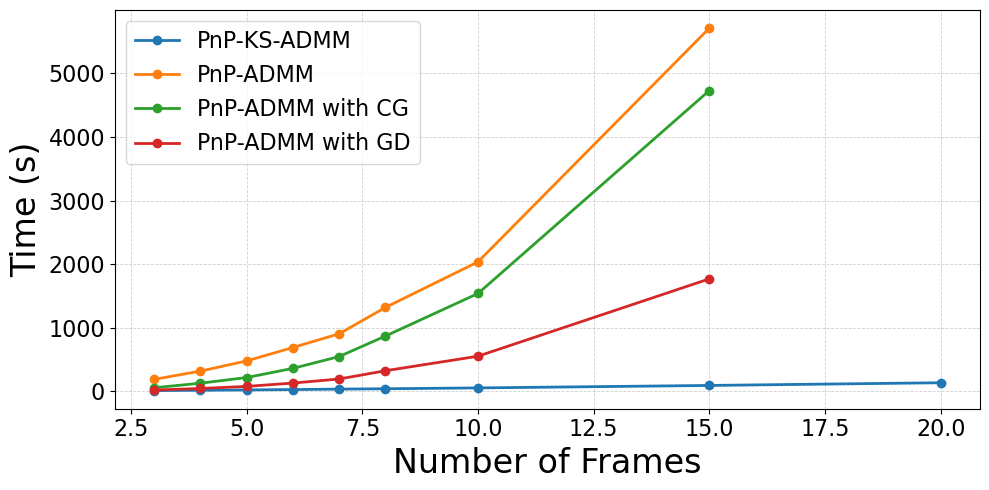} 

    \includegraphics[width=0.38\textwidth]{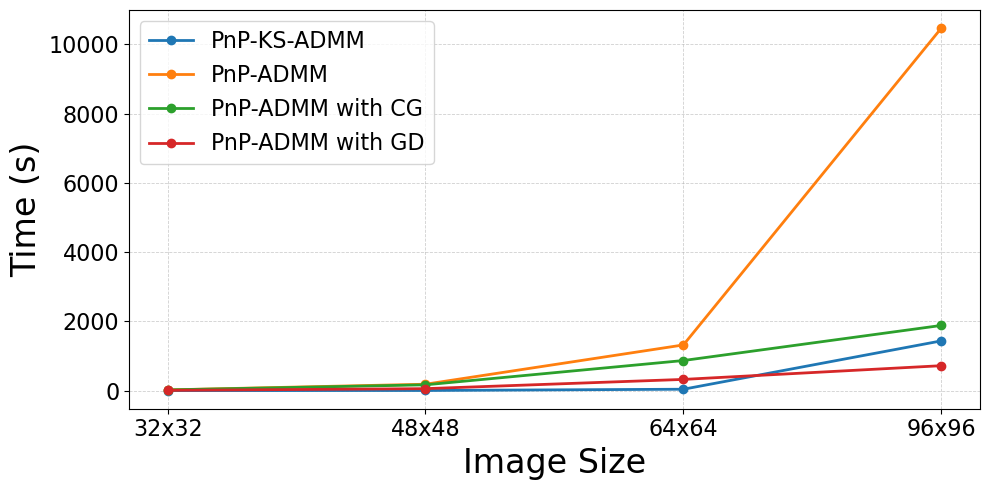} 
    \vspace{-3mm}
    \caption{\small
    Reconstruction time for the considered PnP methods, with respect to (top) frame numbers for fixed image size $N=64^2$, and (bottom) image sizes for fixed frame number $T=5$.}
    \label{fig:frame_time} 
    \vspace{-2mm}
\end{figure}

\begin{figure}[t]
    \centering
    \includegraphics[width=0.4\textwidth]{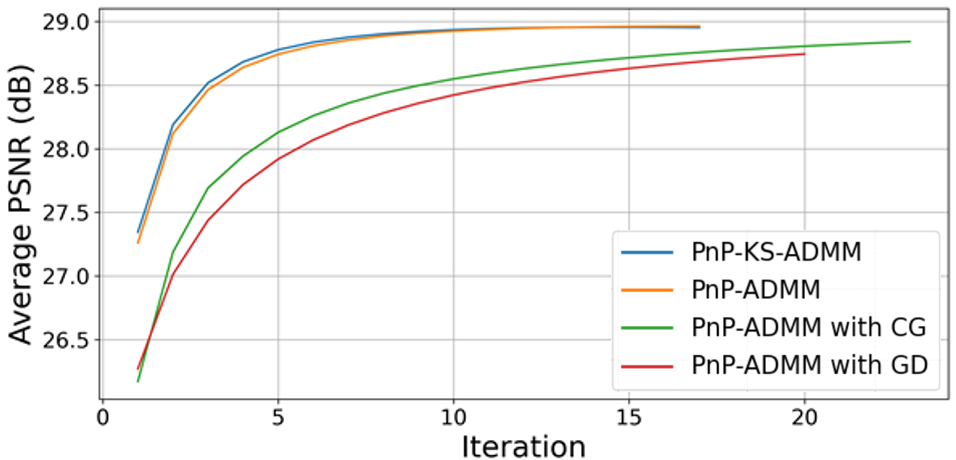} 
    \vspace{-3mm}
    \caption{\small
    Average PSNR profile with respect to iterations for the different considered PnP methods.}
    \label{fig:frame_PSNR}
    \vspace{-6mm}
\end{figure}

The scaling results for increasing $T$ (for $N = 64 \times 64$ fixed) and $N$ (for $T=5$ fixed) are displayed in Figure~\ref{fig:frame_time}. 
For increasing $T$, Figure~\ref{fig:frame_time}(top) shows PnP KS-ADMM scales more efficiently than the other PnP ADMM methods. As a reference, we ran the proposed PnP KS-ADMM when $T=100$ frames, which ran in a time of $1,093s$. This is similar to the time it takes for the PnP ADMM with exact calculation to analyze $T=7$ frames ($907s$). 
For increasing $N$, Figure~\ref{fig:frame_time}(bottom) suggests that the PnP KS-ADMM is more affected by larger image sizes than other methods. Thus, for few frames and large images, standard PnP ADMM may be preferable, while for many frames KS provides higher efficiency.

For the sake of completeness, Figure~\ref{fig:frame_PSNR} shows the PSNR values (averaged over $T$ frames) as a function of iteration $k$. We can observe that PnP KS-ADMM follows the exact reconstruction PSNR more closely than approximations. This shows that the KS achieves higher accuracy across iterations than GD-based approximations.

\section{Conclusion}
\label{sec:Conc}

We have proposed a new algorithm, KS PnP-ADMM, enabling the use of learned regularizer for solving dynamic imaging problems. We shown through simulations on a $2D+t$ deblurring problem, that the KS PnP-ADMM is more computationally efficient than batch solutions for a high number of frames. Future work includes generalization to non-linear dynamic or measurement models. 

\small
\bibliographystyle{IEEEbib}
\bibliography{refs}

\end{document}